\documentclass[AMA,STIX1COL]{WileyNJD-v2}

\usepackage{graphicx}
\usepackage{wrapfig}
\usepackage{pdflscape}
\usepackage{rotating}
\usepackage{epstopdf}
\usepackage{float}
\usepackage{subfig}
\usepackage{amsmath}
\usepackage{bm}
\usepackage{soul}
\articletype{Article Type}

\received{xx}
\revised{x}
\accepted{xx}

\begin{document}

% tikzset{external/force remake}

% NOTES
% refer to Basket Trials and example applicaiton?
% e.g.
% He L, Ren Y, Chen H, Guinn D, Parashar D, Chen C, et al.. Efficiency of a randomized confirmatory basket trial design constrained to control the family wise error rate by indication. Stat Methods Med Res (2022) 31:1207–23. doi:  10.1177/09622802221091901 [PubMed] [CrossRef] [Google Scholar]
%
%

\title{A Bayesian hierarchical mixture cure modelling framework to utilize multiple survival datasets for long-term survivorship estimates: A case study from previously untreated metastatic melanoma}
% \protect\thanks{title footnote.}}

\author[1]{Nathan Green*}

\author[2]{Murat Kurt}

\author[2]{Andriy Moshyk}

\author[3]{James Larkin}

\author[1]{Gianluca Baio}

\authormark{N Green \textsc{et al}}

\address[1]{\orgdiv{Department of Statistical Science}, \orgname{UCL}, \orgaddress{\state{London}, \country{UK}}}

\address[2]{\orgdiv{Worldwide Health Economics and Outcomes Research}, \orgname{Bristol Myers Squibb}, \orgaddress{\state{NJ}, \country{USA}}}

\address[3]{\orgname{The Royal Marsden Hospital}, \orgaddress{\state{London}, \country{UK}}} %james.larkin@rmh.nhs.uk

\corres{*Nathan Green \email{n.green@ucl.ac.uk}}

% \presentaddress{}

% <= 250 words
%ORCID ID: 0000-0002-7056-2741
\abstract[Summary]{
Time to an event of interest over a lifetime is a central measure of the clinical benefit of an intervention used in a health technology assessment (HTA). Within the same trial, multiple end-points may also be considered. For example, overall and progression-free survival time for different drugs in oncology studies. A common challenge is when an intervention is only effective for some proportion of the population who are not clinically identifiable. Therefore, latent group membership as well as separate survival models for identified groups need to be estimated. However, follow-up in trials may be relatively short leading to substantial censoring. We present a general Bayesian hierarchical framework that can handle this complexity by exploiting the similarity of cure fractions between end-points; accounting for the correlation between them and improving the extrapolation beyond the observed data. Assuming exchangeability between cure fractions facilitates the borrowing of information between end-points. We undertake a comprehensive simulation study to evaluate the model performance under different scenarios. We also show the benefits of using our approach with a motivating example, the CheckMate 067 phase 3 trial consisting of patients with metastatic melanoma treated with first line therapy.
}


\jnlcitation{\cname{%
\author{Green N.} et al} (\cyear{2024}), 
\ctitle{}, \cvol{1; 00:1--6}.}

\maketitle

\footnotetext{\textbf{Abbreviations:} MCM, mixture cure model}

\section{Introduction}\label{sec:intro}
%%% HTA perspective %%%
% survival data
Interventions that impact the time to an event of interest, such as disease progression or death in a cancer trial, form a high proportion of appraisals by Health Technology Assessment (HTA) agencies, such as the National Institute for Health and Care Excellence (NICE), in England \citep{Latimer2011}.
For instance, a new intervention may be expected to increase the amount of time until a patient experiences disease progression or death from any cause. For this reason, ``survival data'' are instrumental to modelling in HTA \cite{Demiris2006, Jackson2010}.

% what is a data-cut?
A feature of survival data from a trial is that they are often reported on during as well as at the end of the trial period.
This enables ongoing assessments and can inform decisions such as an early end to the trial due to harm or futility.
However, in reality, it is normally not possible to make use of the collected data in real time because there is a period in which they need to be cleaned due to issues, such as missingness and other data quality errors.
Therefore, snapshots of the trial data are made at pre-specified points defined in the study protocol.
This can be on a particular date, follow-up time, or a particular number of patients.
These subsets of survival data made at discrete times are called \textit{data-cuts}.

% extrapolation
In order to quantify accurately the health and economic benefits of a new intervention using such survival data, it is necessary to estimate the {\it mean survival time}, ie,~the \textit{long-term} effects of a given intervention.
This is in contrast to general survival curve summaries, such as the median time, which are typically used in standard ``biostatistical'' analyses.
This idiosyncrasy has important implications because, in order to calculate the mean survival time we require the full survival curve, ie, over a patient's lifetime, but available data (e.g.~data-cuts from a randomised trial) almost inevitably only cover a limited time frame and are subject to a high degree of censoring.
Thus, it is necessary to \textit{extrapolate} the observed survival curve to a time horizon that is typically much longer than the one considered in the experimental study.
Consequently, unlike ``standard'' time-to-event analyses that are often based on semi-parametric models (e.g.~Cox regression), HTA modelling is based on a fully parametric approach to the survival analysis, as recommended by a highly influential NICE Decision Support Unit (DSU) Technical Support Document (TSD) \citep{Latimer2011}.

% plateaus
Another important feature of survival data in the field of cancer research has arisen in the past decade due to the development of potentially highly successful immuno-oncology treatments.
These aim to stimulate the body’s immune system to recognise and kill cancer cells \cite{Ouwens2019}. Several types of immunotherapy can be used to treat melanoma including immune checkpoint inhibitors, interleukin-2 (IL-2), and oncolytic virus therapy.
In particular, inhibitory immune checkpoint blockade combination therapies have shown significant success in promoting immune responses against cancer and can result in tumour regression in many patients \cite{Khair2019}.
These advances have lead to cancer patients with improved survival end-points and more long-term survival (LTS) (although complete responders, CR, still have worse outcomes than the general population).
%%TODO: what do we mean by this (vs. the general population)? CR patients do the best vs the patients in the study who do not achieve a CR, but are we comparing the CR patients to the real world melanoma population regardless of response to a given therapy? 
In previous immuno-oncologic studies for melanoma therapies, such as those evaluating ipilimumab and nivolumab, results have indicated that survival curves ``plateau'' for a considerable proportion of LTS patients in the time period covered by survival curve \citep{Wolchok2017, Larkin2019}.
That is, the underlying survival curve converges to a probability (substantially) greater than zero and does not appear to decrease further.
Note that in practice, in the tail of the Kaplan-Meier plateauing may also be due to sampling issues, such as small sample size or missing data, and so there will be uncertainty about the true proportion of LTS.
% mixture cure models
When a survival curve exhibits this plateau behaviour, it is common to adopt a \textit{mixture cure model} (MCM) approach to describe the generation of the underlying data. A MCM considers a population as a mixture of two groups: the cured and not cured. In our case, the cured group contains the LTS patients. In many MCM analyses the survival associated with the uncured fraction of the population is represented by a Cox proportional hazard model.
% excess hazard models
Alongside this consideration, it is also important to account for the additional risk experienced by the uncured patients due to the cancer, beyond the background mortality risk of the general population (the cured). \textit{Excess hazard} models provide a framework to separate the baseline mortality (background hazard) from the additional risk associated with the disease. This can be especially important when dealing with survival data where patients may not all have the same risk of death.

% data-cut extrapolation 
However, the survival plateaus indicating a proportion of LTS may not have been reached at the time of a data-cut (including the planned end of the study).
In order to obtain complete survival curves to use in an HTA evaluation we need to extrapolate the survival curves to the sustained plateaus.
Ideally, these extrapolations should be consistent between data-cuts \citep{Bullement2020}.
The question of how to principally and usefully do this in situations where trials include multiple treatment arms and end-points is the subject of this paper.
The fact that we consider multiple end-points within a trial in the same analysis means that information about the plateau from data for one end-point can potentially help inform the estimation of the plateau for another end-point.
This can enable consistency between plateau estimates and maximise the utility of all available data.

% review of related work
Within a frequentist paradigm, an example of related work is a multi-level modelling approach in mixture cure modelling using random effects to model multilevel clustering structure in the linear predictors in both the hazard function and cured probability parts \cite{Lai2009}.
Correlation between random effects in the uncured survival and cure fraction has been investigated using a bivariate Normal distribution \cite{Lai2008} \cite{Tan2018}.
Instead, we will impose dependency with hyperparameters.
% Tan2018 model log HR rather than cure fraction
To our knowledge there has been no Bayesian fully-parametric mixture cure model with a multi-level modelling structure for end-points.

% plan of the paper
The paper is structured as follows.
In the next section, the motivating example and data used throughout is outlined.
We introduce the basic mixture cure model in Section~\ref{sec:basic_model}, before extending this to our Bayesian hierarchical model in Section~\ref{sec:hier_model}.
In Section~\ref{sec:application} our novel modelling approach is applied to the motivating example dataset.
Comparisons are made between our approach and an independent model analogue.
Finally, in Section~\ref{sec:discussion} we discuss the results and future work.

\section{Motivating example}\label{sec:example}
Our motivating example concerns a long-term study of melanoma therapies.
The study has been described in detail elsewhere, so we give a brief overview of the data here (see \citep{Wolchok2017, Larkin2019, Hodi2018} for more details).
The \textit{CheckMate 067} trial is a phase 3 randomised, double-blind study conducted on eligible patients aged 18 years or older with previously untreated, unresectable stage III or IV melanoma.
Nivolumab alone or the combination of nivolumab plus ipilimumab was compared with ipilimumab alone in patients with metastatic melanoma.
They were randomly assigned in a 1:1:1 ratio and stratified by anti–programmed death-1 ligand 1 (PD-L1) expression status, whether there is a mutation in the BRAF gene that makes it work incorrectly (BRAF mutation), and AJCC metastasis stage.
% The purpose of the trial was to test the effectiveness of ipilimumab, an anti–cytotoxic T-lymphocyte–associated antigen 4 monoclonal antibody, and nivolumab, an anti–programmed death 1 agent.
The total baseline sample consists of $n = 945$ subjects divided across three treatment arms
i) ipilimumab monotherapy with 3 mg/kg IV once every 3 weeks (Q3W) for a total of 4 doses;
ii) nivolumab monotherapy with 3 mg/kg intravenous (IV) once every 2 weeks (Q2W);
iii) combined nivolumab with ipilimumab with 1 mg/kg IV and 3 mg/kg IV Q3W, respectively, for 4 doses followed by nivolumab 3 mg/kg IV Q2W.
The co-primary end-points were \textit{progression-free survival} (PFS; ie,~the time from randomisation until the first of disease progression or death) and \textit{overall survival} (OS; ie,~the time from randomisation to time of death).
Disease progression was defined as evaluated tumour response using criteria from RECIST (Response Evaluation Criteria in Solid Tumors), version 1.1. This includes a significant increase in size, new lesions or worsening of existing lesions.
The data source for the analysis consists of patient-level data of times to PFS and OS with covariates sex, age at the start of the trial, and country of centre.

In our data set the minimum study follow-up from randomisation of 60 months was used as the final data-cut time.
We also considered two earlier artificial data-cuts by truncating the 60-month data at 12 and 30 months.
These particular times were chosen for this analysis because the actual study had a planned data-cut at 28 months, and the largest observed median time for OS was just over 30 months. Thus, this time is reasonable and relevant to the trial, and at this time it may be possible to make comparable inferences to the complete 60-months data set but at a significantly earlier time.
Moreover, the latest median time between all treatment regimens for PFS was approximately 12 months. Therefore, at 12 months it may be possible to characterise the PFS survival curves but not the less mature OS and so borrowing information between PFS and OS may be advantageous. Taking a data-cut at an even earlier time would probably not provide sufficient information to enable any useful extrapolation.
Patients who were still being followed-up but had not experienced the event of interest at the time of an artificial data-cut were censored at this time.

\begin{figure}[!ht]
\centering
\includegraphics[width=0.6\linewidth]{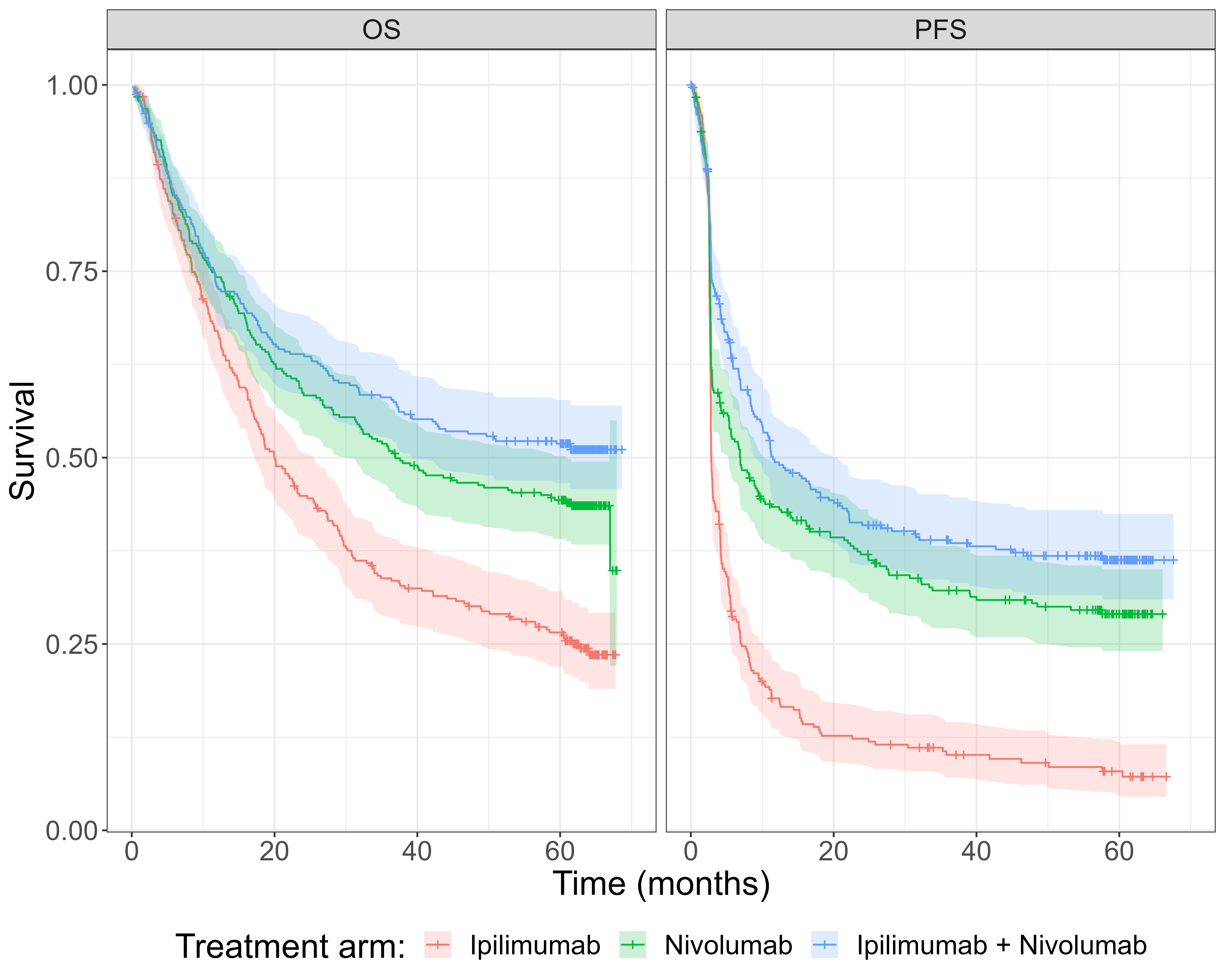}
\caption{\label{fig:S_raw_data} Kaplan-Meier curves of PFS and OS times with 95\% CI for the complete CheckMate 067 phase 3 trial data and ipilimumab, nivolumab and combination treatments.}
\end{figure}

The reader is encouraged to refer to previously published analyses for the CheckMate 067 trial results, including the 5-year database lock (DBL) publication\cite{Larkin2019}. In summary, for PFS the median (95\% CI) times to event for each treatment arm were 2.86 months (2.79-3.29) for ipilimumab, 6.93 months (5.32-10.41) for nivolumab, and 11.50 months (9.26-20.80) for the combination treatment of nivolumab and ipilimumab. For OS the median (95\% CI) times to event for each treatment arm were 20.0 months (17.22-25.6) for ipilimumab, 36.9 months (31.24-60.9) for nivolumab, and not observed for the combination treatment.
Figure~\ref{fig:S_raw_data} shows the Kaplan-Meier estimates of the PFS and OS survival curves for the complete data from the CheckMate 067 phase 3 trial. Both sets of curves, for PFS and OS, appear to approach a positive limiting probability; the PFS curves more clearly so. Early published results for outcomes at 7.5 years follow-up continue to demonstrate the durability of responses with combined nivolumab and ipilimumab, and an ongoing survival plateau~\cite{Hodi2022}.

A standard frequentist analysis has been carried-out on these data already with separate mixture cure models for each treatment and OS or PFS \cite{Mohr2020}.
It was estimated that for OS, ranges of estimated proportions of LTSs were 16\%–26\% for ipilimumab, 38\%–46\% for nivolumab, and 49\%–54\% for combined treatment across all modelled distributions.
Similarly, for PFS, ranges of the proportions of LTSs were 9\%–13\% for ipilimumab, 29\%–33\% for nivolumab, and 38\%–40\% for combined treatment.

\section{Modelling framework}\label{sec:methods}
Consider a study in which $n$ individuals are observed, and the (non-negative) time to either progression ($T^{prog}_{i}$) or death ($T^{death}_{i}$) is recorded for each individual $i=1,\ldots,n$.
Denote the event time for individual $i$ by $T_{i}$.
In our motivating example, we would have a separate set of $T_{i}$ for either end-point, PFS or OS.
Thus, PFS times are $T_{i} = \min(T^{prog}_{i}, T^{death}_{i})$, and OS times are $T_{i} = T^{death}_{i}$.
Event times may be censored, and therefore we consider a censoring time $C_{i}$ and an event indicator $\delta_{i} = \mathbb{I}(T_{i} < C_{i})$
--- so that $\delta_{i} = 1$ if the end-point is fully observed for individual $i$ and 0 if it is censored.
The observed survival time is then indicated as $t_{i} = \min(T_{i}, C_{i})$.

Typically, a study will also measure some individual-level covariates, which we assume are collected in a vector $\bm{x}_{i} = (x_{1i}, \ldots, x_{Pi})^\top$.
These may include the individual's age, sex, comorbidities, and usually, a treatment arm indicator.
The complete data can then be represented as
$\mathcal{D}_i = (t_i, \delta_i, \bm{x}_i)$.
% $\mathcal{D}_i = (\bm{t}_i, \bm{\delta}_i, \bm{x}_i)$.
% where $\bm{t}_i = (\bm{t}_{1i}, \bm{t}_{2i})$, $\bm{x}_i = (\bm{x}_{1i}, \bm{x}_{2i})$ and $\bm{\delta}_i = (\bm{\delta}_{1i}, \bm\delta_{2i})$.
In our example, the covariates are the same between PFS and OS.

We modelled the observed times using a parametric sampling distribution $t_{i} \sim p(t_{i} \mid \bm\theta)$ as a function of a set of model parameters $\bm\theta = (\bm\alpha(\bm x), \mu(\bm x))$. 
In this notation, the location parameter $\mu(x)$ indicates the mean or scale of the probability distribution; and a (set of) ancillary parameters $\bm \alpha(x)$ describe the shape or variance of the distribution.
Although it is possible for both $\mu$ and $\bm\alpha$ to explicitly depend on the covariates $\bm x$, usually the formulation is simplified to assume that these only directly affect the location parameter.
Since $t>0$, we typically use a generalised linear formulation.
So, for an individual $i$
$$
g(\mu) = \beta^{\mu}_{0} + \sum_{p=1}^P \beta^{\mu}_{p} x_{pi},
$$
to model the rate or location parameter,
where $\beta^{\mu}_{0}$ represents an intercept,
and $\beta^{\mu}_{p}$ are the coefficients for an individual's measurements.
The formula may also include additional components for the linear predictor (e.g.~structured random effects).
The function $g(\cdot)$ is typically the logarithm.

\subsection{The standard mixture cure model} \label{sec:basic_model}
Suppose that we observe a sustained plateau in the survival time data or have some other reason to believe that there is a long-term survivor group in our sample. We will now present the standard mixture cure model (sometimes called a long-term survival model) as the basis of our proposed extension.
Denote the probability of being cured by the {\it cure fraction}, $\pi$.
The term "cure" here refers to the type of statistical model and does not literally mean that they are cured of disease. In our case, cured means that a patient has responded to treatment and is a LTS.
Because the cure models explicitly split patients into two groups (cured or not) then the full set of model parameters consists of two separate groups. Define $\bm\omega = (\bm\theta^b, \bm\theta^u)$, where superscript $b$ denotes the \textit{background} model and $u$ denotes the \textit{uncured} model. In the standard cure model, cured individuals are usually assumed to have no chance of experiencing the event of interest in the period under consideration, simplifying the model. There are several types of cure models and a rich and growing body of research in this area. (see \citep{Yu2013} for a comparison and guidance with application to oncology.)

For an individual $i$ in a population with a cure fraction, the mixture cure model is associated with the following survival function

\begin{equation}
\label{eqn:mcm}
S(t_{i} \mid \bm\omega, \bm{x}_i) = S_b(t_{i} \mid \bm\theta^b, \bm{x}_{i}) \left[\pi + (1 - \pi) S_u(t_{i} \mid \bm\theta^u, \bm{x}_{i}) \right],
\end{equation}
\\
\noindent
where $S(t) = 1 \!-\! \int_0^t p(s \mid \theta)\, \text{d}s$ denotes survival at time $t$,
$S_b(t \mid \bm\theta^b, \bm{x})$ is a function of the background mortality at time $t$ conditional on covariates $\bm{x}$,
% some papers denote \pi as _uncured_ fraction
and $S_u(t \mid \bm\theta^u, \bm{x})$ is a function of the (excess) mortality due to cancer at time $t$, conditional on covariates $\bm{x}$. The formulation in equation~(\ref{eqn:mcm}) can be thought of as comprising of two submodels: i) an \textit{incidence} model for the cure fraction $\pi$ (the general term from epidemiology is the rate or probability of a disease); and ii) a \textit{latency} model for the survival function of the uncured population $S_u$. This is `hidden' since it cannot be directly observed from the data.

The model above can be extended to account for multiple treatments, $K$; for instance, in our running example, $K=3$ and we could specify the cure fraction for treatment $k=1,\ldots,K$ using the following ``fixed'' effect model
\begin{equation}
\label{eqn:pi_regn}
\mbox{logit}(\pi_{k}) = \beta^{\pi}_{k},
\end{equation}
\noindent
where $\beta^{\pi}_k$ are the regression coefficients quantifying the impact of treatment $k$.
The formula could also contain additional terms, such as a frailty term.
The treatment index for an individual is included in the covariate vector $\bm{x}$ so that $\pi$ is replaced by $\pi(\bm{x})$ in equation~(\ref{eqn:mcm}).
% We will assume without loss of generalisability that all individuals have all treatments.
Note that although we are including all treatment arms in the same model we are not assuming any relationship between them. In particular, we do not suppose proportional hazards. This allows greater flexibility for the cases when the proportional hazards assumption is violated.

\subsubsection{Issues modelling multiple end-points} \label{sec:issues}
There are several potential issues with existing approaches to cure modelling, when applied to an HTA context, in particular.

% correlated events
Firstly, when the end-points are correlated, such as with the common events of interest overall survival (OS) and progression-free survival (PFS), this ought to be accounted for in the modelling. There may be information in one sample of event times that we can use to improve the inference from another. Performing separate, independent analyses may even give counter-intuitive results. In the case of multiple types of end-points in the same trial, since the clinical interpretation of a cure fraction is the proportion of patients that are LTS, it may make intuitive sense that there is a single underlying cure fraction. For example, what would be the interpretation if there are differences in emergent survival plateaus between PFS and OS? This is a clinically unintuitive dichotomy between the resulting proportions of long-term survivors.

% censoring
Secondly, time-to-event data are typically right-censored due to administrative censoring or loss to follow-up.
The censoring times may be too early in order to adequately characterise a survival model for HTA for a single end-point. In particular, OS times are bounded below by PFS times by definition, so will be subject to more censoring at the same cut-point. By taking advantage of the set of end-points together, the data paucity may be alleviated to give better fits and extrapolation to the full lifetimes of patients. This is particularly important for health economic modelling to gauge the full cost and effect of an intervention.

% complexity penalising
Furthermore, if there is heavy censoring for all endpoints, such as early data cuts, and the number of endpoints is small, then the data alone may not represent the true underlying survival curves, including the plateau for the cure fractions.
At this point, additional sources of information should be principally incorporated into the model. A Bayesian paradigm naturally allows for the inclusion of external evidence from sources such as historical data and expert opinion. It is reasonable to assume, in the absence of additional information, that the cure fractions for all end-points are drawn from the same or very similar distribution. In order to do this, we can increase the amount of strength to borrow between end-point data which strikes a balance between the information (or lack of) in the data and other knowledge \cite{Nikolaidis2021}. Practically, this corresponds to a reduction in the global cure fraction variance hyper-parameter which in the limit is a base model with $\sigma_k=0$, equivalent to complete pooling of the data. 

% our innovative proposal
We will now present a Bayesian hierarchical mixture cure model for multiple end-points.
This has the double advantage of borrowing information between end-points
(e.g., in the case of OS and PFS, the likely more mature PFS data can inform the often highly censored OS analysis) and obtaining a single cure fraction estimate.

\subsection{The hierarchical mixture cure model} \label{sec:hier_model}
In this section, we show how to construct the extended mixture cure model in order to model multiple end-points together and alleviate the issues posed in the Section \ref{sec:issues}. In order to do this, we must first introduce an additional index to denote an end-point $j = 1, \ldots, J$ --- typically, $J=2$ for PFS and OS. We can now define a separate mixture cure model, as described in equation~(\ref{eqn:mcm}), for each of the end-points $j$. With this in mind, let us consider three alternative multiple end-point incidence models for the estimation of the cure fraction for treatment $k$ and end-point $j$, denoted $\pi_{kj}$.

First, the approach taken so far, and presented in Section \ref{sec:basic_model}, is to model the cure fraction corresponding to each end-point completely separately (no pooling), assuming that they are independent.
$$
\pi_{kj} \perp\!\!\!\perp \pi_{kj'}, \; j,j' = 1, \ldots, J.
%\text{logit}(\pi_k) \sim \text{Normal}(\mu_k, \sigma_k^2), \; k = 1, \ldots, K.
$$
Secondly, we can assume that the cure fraction is the same for all end-points and so pool (or "lump" together) the samples (complete pooling). In our example, PFS can be used as a proxy for OS since there are fewer missing data.
There may be no reason why the cure fraction should be different between different end-points, especially if individuals are observed for long enough. In this case
$$
\pi_{kj} = \pi_{kj'}, \; j,j' = 1, \ldots, J.
$$
A potential issue with the complete pooling approach is that it ignores any variation in cure fraction that exists between end-points.

\begin{figure}[!ht]
\centering
\includegraphics[height=7cm, width=0.55\linewidth]{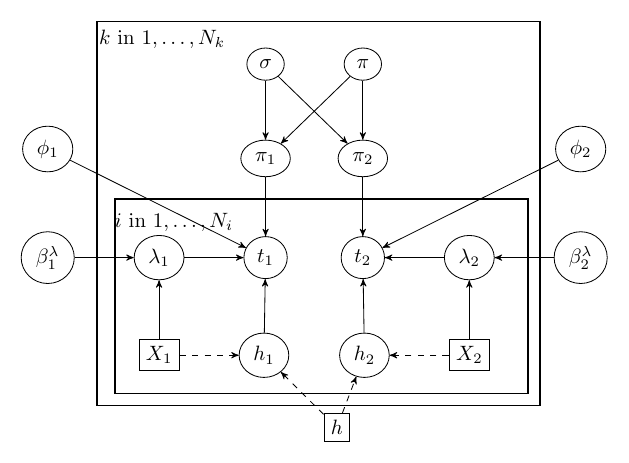}
\caption{\label{fig:hier_dag} A hierarchical mixture cure model DAG for a trial with two end-points. The trial indices correspond to the motivating example end-points of PFS (1) and OS (2).
Solid lines represent stochastic and dashed lines deterministic relationships, respectively.
Cured patients have fixed hazards, e.g. taken from life tables.
The distribution of times for uncured patient regresses on covariates for the rate parameter and $\phi$ is the set of ancillary parameters.}
\end{figure}

Lastly, and the approach proposed in this paper, is a compromise between the first two approaches called partial-pooling.
We propose a hierarchical structure on the cure fraction assuming exchangeability between all end-points $j$ for each treatment $k$.
We assume that there is a single, shared "common" cure fraction from which the separate end-point specific cure fractions are obtained, corresponding to equation~(\ref{eqn:pi_regn})
but where $\pi_{k}, \beta^{\pi}_{k}$ are no longer for particular end-points as before but now represent "global" values.
Using the global cure fraction, we can define the second-level end-point specific cure fractions as

\begin{equation}\label{eqn:global_cf}
\text{logit}(\pi_{kj}) \sim \text{Normal}(\nu_k, \sigma_k^2), \; j = 1, \ldots, J.  
\end{equation}
\noindent
where $\nu_k = \text{logit}(\pi_k)$ and $\sigma_k^2$ is the between-group (end-points) variance.
The end-point specific cure fractions in our example are denoted $\pi_{PFS}, \pi_{OS}$.
Figure~\ref{fig:hier_dag} shows a directed acyclic graphical (DAG) representation of the Bayesian hierarchical mixture cure modelling framework for two end-points.
The equivalent DAGs for the complete and no pooling models are given in the Appendix.
Note that pooling refers to the cure fractions. The survival times are not pooled; the parameters for the uncured survival models ($\phi$, $\beta^{\lambda}$ in the latency model) are estimated using the separate sets of survival times for each end-point e.g., in the complete pooling case we do not assume that PFS and OS have the same survival time distributions. However, in the complete and partial pooling case, there is an indirect pooling effect for the latency models via the incidence model, so the fact that the cure fractions are similar/the same will influence the shapes of the survival curves. A more general DAG for any number of end-points would have an additional plate (box) around $\pi_{jk}$ with a single individual level model nested within it.

\section{Simulation study}\label{sec:simulation}
To assess the performance of our method in addressing the issues set out above, we conducted a simulation study and compared the outcome between defined scenarios \cite{Burton2006}. We followed the structured approach for designing and reporting simulation studies, described in \cite{Morris2019}, which involves defining aims, data-generating mechanisms, estimands, methods, and performance measures (“ADEMP”).
% The conducted simulation study was motivated by the CheckMate067 trial.

\subsection{Scenarios and data generating process}
We simulated survival times from various hierarchical mixture cure models. We decided to do this rather than resampling from an existing data set because of the flexibility and extended scenarios this made available. For example, the real data example in Section~\ref{sec:application} would be constrained to only two end-points.

% target statistics
The target estimates used were the restricted mean survival time (RMST) and cure fractions ($\pi$). The median survival time ($t_{0.5}$) was not used. These were chosen because they are particularly relevant for health economic evaluation. An alternative approach would have been to use, e.g. the Concordance Index (C-index) or log-rank test more favoured in the field of survival analysis. We shall use a particular version of RMST appropriate for a mixture cure model\footnote{In a similar way, the median survival time would be adjusted for the cure fraction model. Where the median time for the uncured fraction is that $t$ such that $S_u(t) = 0.5$, for the entire population, we require $S(t) = 0.5$. If we assume that there is no additional background mortality, then $t_{0.5}$ is the solution of $S_u(t) = (0.5 - \pi) / (1 - \pi)$.}. Defining $\tau$ as the maximum (cut-off) time then
$$
RMST = \pi \tau + (1-\pi) \sum_0^{\tau} S(t)\; \text{d}t.
$$

The underlying data generating model is of the form
\begin{align}
    \notag
    \text{logit}(\pi_j) &\sim \text{N}(\mu_{\text{true}}, \sigma^2_{\text{true}}) \\
    \notag
    s_{ij} &\sim \text{Bern}(\pi_j)\\
    \label{eqn:data-generation}
    t^{u}_{ij} &\sim \text{Weibull}(k_j, \lambda_j)\\
    \notag
    t_{ij}^{\text{obs}} &= 
    \notag
    \begin{cases} 
    \notag
    t^{u}_{ij}, & \text{if } s_{ij} = 0 \text{ (not cured)} \\
    \notag
    \tau, & \text{if } s_{ij} = 1 \text{ (cured)},
    \notag
    \end{cases}
    \notag
\end{align}
% include random censoring
where event time $t_{ij}^{\text{obs}}$ for individual $i$ and end-point $j$ are generated from a Weibull distribution with the only censoring at time $\tau$ due to being cured. No covariates were included. An individual's cured status $s_{ij}$ is generated from a Bernoulli distribution with probability $\pi_j$ sampled from a global cure fraction Normal distribution on the logit linear scale with parameters $\mu_{\text{true}}, \sigma^2_{\text{true}}$. We don't include any censoring in the data generation because this has a similar effect to sample size for our purposes and would further increase the number of simulation scenarios. That is, introducing censoring is a form of information loss which reduces the effective sample size (although still providing partial information). 

% performance measures
The performance measures calculated were bias, empirical standard error (empSE) and coverage.
Bias is the difference between the expected value of an estimator and the true value. The empSE is the true variability of the estimator. Finally, the coverage probability is the probability that the credible intervals contain the true value which we use the 95\% level.
We also included relative bias (RB), which is the bias scaled by the true value, because different values of $\pi_j$ are used for different data-generating mechanisms and this permits a more straightforward comparison.
Target mean estimates $\widehat{RMST}$ and $\widehat{\pi}$ were obtained from posterior samples.

% true values
The underlying true values that were held fixed between scenarios were selected as follows. The between-group cure fraction standard deviation was held fixed because the simulation study is concerned with the relation with the cure fraction priors. We also make the assumption that it is narrow enough to justify a hierarchical model and exchangeability. Note again that $n_{sim}$ can be thought of in a sense as an equivalent larger sample size in the presence of censoring.
The true global cure fraction was 0.2; the global cure fraction standard deviation was assigned a zero-truncated Cauchy of $\text{Cauchy}^{+}(0.4, 2.4)$ (corresponding to mean 19, median 2.4 and 95\% interval 0.1 - 55);
for the separate cure fraction model, the cure fraction on the logistic scale was assigned the prior distribution $\text{N}(-1, (0.2 + 0.4)^2)$ to correspond to the hierarchical specification;
cut-point time $\tau = {5}$; the true latent survival model distributions were a 1:1 ratio of $\text{Weibull}(k=1, \lambda=1)$ and $\text{Weibull}(k=1, \lambda=4)$. This was chosen to ensure that half of the curves had not converged sufficiently to the plateau at the cure fraction, and thus allowed for the potential borrowing of information.

% scenarios
The factors to vary were chosen over two levels of model inputs, corresponding to high or low values. These factors were varied fully factorially. For the purposes of clarity and space, a selection of these will be presented here, and the full set of plots and tables is presented in the Appendix. We shall reference items in the Appendix with a preceding A. In total, the analysis consists of 32 scenarios as follows.

% describe scenarios
In terms of the data simulation, the endpoints, i.e. number of survival curves, was $n_e = {3,10}$, and sample size was $n_{sim} = {10,100}$. In terms of the model prior distributions, the latent survival curves were given informative priors ${k \sim \text{Gamma}(1000, 1000)}$, and either $\lambda \sim \text{logN}(1.4, 0.002)$ or $\lambda \sim \text{logN}(0, 0.01)$; and weak priors $k \sim \text{Gamma}(1000, 100)$, and either $\lambda \sim \text{logN}(1.4, 0.02)$ or $\lambda \sim \text{logN}(0, 0.1)$. The former corresponds to the situation when we have good information about the underlying (uncured) curves but perhaps not for the cure fractions.
Similarly, the global cure fraction scenarios are for when the global mean is strongly centred at the true value or weak centred at 0.5 [95\% interval: 0.2 - 0.8], indicating little prior knowledge; the mean global cure fraction priors on the logit scale were either informative $\beta^{\pi} \sim N(-1, 0.01)$ or weak $\beta^{\pi} \sim N(0, 0.7)$. Finally, the between-group cure fraction prior standard deviation was assigned an informative prior $\sigma \sim \text{logN}(0.05, 0.05)$ and a weak prior $\sigma \sim \text{logN}(0.05, 0.1)$. The full list of scenarios is presented in Table~A18 in the Appendix.
Finally, note that the varying of the different prior distributions is a type of robustness analysis, from which we can investigate the influence of the different choices. Having the global cure fraction prior centred at 0.5 in our analysis is an example of what's called an `optimistic' prior, i.e. bias upward.

\subsection{Simulation results}
The main analysis produced 1000 posterior samples for each scenario data set, which were replicated 1000 times following equation~(\ref{eqn:data-generation}). The runtime exceeded several hours using a Myriad high-performance cluster. Additional figures and tables of results found in the Appendix include histograms of posterior distributions of the performance measures (Figures~A18 and A19), lollipop plots comparing separate and hierarchical results (Figures~A20-A25), and zip plots of coverage (Figures~A26-A34). For an overall summary, Table~\ref{tab:performance_measures} shows the performance measure estimates averaged over all curves within scenarios. Figures A22-23 in the Appendix give corresponding heat maps for easy pattern recognition.

% That is, $\overline{\widehat{\pi}} = \sum_j \widehat{\pi}_j$, $\overline{\widehat{RMST}} = \sum_j \widehat{RMST}_j$, $\overline{\widehat{RB}} = \sum_j \widehat{RB}_j$ and $\overline{\widehat{t_{0.5}}} = \sum_j \widehat{t_{0.5}}_j$.

\begin{table}[h!]
\centering
\begin{tabular}{rrrrrrrrrrrrrrrrr}
\toprule
\multicolumn{1}{c}{ } & \multicolumn{8}{c}{RMST} & \multicolumn{8}{c}{Cure Fraction} \\
\cmidrule(l{3pt}r{3pt}){2-9} \cmidrule(l{3pt}r{3pt}){10-17}
\multicolumn{1}{c}{ } & \multicolumn{2}{c}{Bias} & \multicolumn{2}{c}{Coverage} & \multicolumn{2}{c}{empSE} & \multicolumn{2}{c}{RB} & \multicolumn{2}{c}{Bias} & \multicolumn{2}{c}{Coverage} & \multicolumn{2}{c}{empSE} & \multicolumn{2}{c}{RB} \\
\cmidrule(l{3pt}r{3pt}){2-3} \cmidrule(l{3pt}r{3pt}){4-5} \cmidrule(l{3pt}r{3pt}){6-7} \cmidrule(l{3pt}r{3pt}){8-9} \cmidrule(l{3pt}r{3pt}){10-11} \cmidrule(l{3pt}r{3pt}){12-13} \cmidrule(l{3pt}r{3pt}){14-15} \cmidrule(l{3pt}r{3pt}){16-17}
Scenario & Hier & Sep & Hier & Sep & Hier & Sep & Hier & Sep & Hier & Sep & Hier & Sep & Hier & Sep & Hier & Sep\\
\midrule
1 & 0.01 & 0.01 & 0.99 & 0.86 & 0.03 & 0.00 & 0.00 & 0.00 & 0.00 & 0.00 & 0.99 & 0.84 & 0.01 & 0.00 & 0.01 & 0.00\\
2 & 0.01 & 0.01 & 0.98 & 0.85 & 0.03 & 0.00 & 0.00 & 0.00 & 0.00 & 0.00 & 0.98 & 0.83 & 0.01 & 0.00 & 0.01 & 0.00\\
3 & 0.01 & 0.01 & 0.96 & 0.84 & 0.03 & 0.01 & 0.00 & 0.00 & 0.00 & 0.00 & 0.96 & 0.84 & 0.01 & 0.00 & 0.00 & 0.00\\
4 & 0.01 & 0.01 & 0.94 & 0.85 & 0.03 & 0.01 & 0.00 & 0.00 & 0.00 & 0.00 & 0.93 & 0.84 & 0.01 & 0.00 & 0.00 & 0.00\\
\addlinespace
5 & 0.57 & 0.50 & 0.21 & 0.10 & 0.22 & 0.16 & 0.18 & 0.16 & 0.04 & 0.00 & 0.95 & 0.83 & 0.08 & 0.00 & 0.15 & 0.01\\
6 & 0.55 & 0.42 & 0.34 & 0.14 & 0.28 & 0.21 & 0.19 & 0.14 & 0.06 & 0.00 & 0.94 & 0.83 & 0.09 & 0.00 & 0.22 & 0.01\\
7 & 0.57 & 0.39 & 0.12 & 0.02 & 0.14 & 0.09 & 0.19 & 0.14 & 0.12 & 0.01 & 0.35 & 0.72 & 0.05 & 0.01 & 0.46 & 0.05\\
8 & 0.52 & 0.37 & 0.18 & 0.04 & 0.15 & 0.10 & 0.18 & 0.14 & 0.10 & 0.01 & 0.48 & 0.75 & 0.05 & 0.00 & 0.36 & 0.04\\
\addlinespace
9 & 0.01 & 0.01 & 1.00 & 0.97 & 0.06 & 0.01 & 0.00 & 0.00 & 0.00 & 0.00 & 1.00 & 0.97 & 0.02 & 0.00 & 0.01 & 0.00\\
10 & 0.01 & 0.01 & 0.99 & 0.97 & 0.06 & 0.01 & 0.01 & 0.00 & 0.00 & 0.00 & 0.99 & 0.97 & 0.02 & 0.00 & 0.01 & 0.01\\
11 & 0.01 & 0.01 & 0.98 & 0.97 & 0.04 & 0.02 & 0.00 & 0.00 & 0.00 & 0.00 & 0.98 & 0.97 & 0.02 & 0.01 & 0.00 & 0.00\\
12 & 0.01 & 0.01 & 0.97 & 0.97 & 0.04 & 0.03 & 0.00 & 0.00 & 0.00 & 0.00 & 0.97 & 0.97 & 0.01 & 0.01 & 0.00 & 0.00\\
\addlinespace
13 & 0.61 & 0.51 & 0.22 & 0.13 & 0.23 & 0.16 & 0.20 & 0.16 & 0.06 & 0.00 & 0.96 & 0.97 & 0.09 & 0.00 & 0.21 & 0.02\\
14 & 0.57 & 0.43 & 0.37 & 0.20 & 0.27 & 0.22 & 0.20 & 0.14 & 0.07 & 0.00 & 0.94 & 0.97 & 0.09 & 0.00 & 0.27 & 0.02\\
15 & 0.57 & 0.42 & 0.12 & 0.04 & 0.14 & 0.09 & 0.19 & 0.14 & 0.12 & 0.03 & 0.35 & 0.65 & 0.05 & 0.01 & 0.47 & 0.11\\
16 & 0.51 & 0.39 & 0.18 & 0.06 & 0.15 & 0.10 & 0.18 & 0.15 & 0.10 & 0.02 & 0.48 & 0.73 & 0.05 & 0.01 & 0.36 & 0.09\\
\addlinespace
17 & 0.20 & 0.36 & 0.93 & 0.92 & 0.19 & 0.20 & 0.08 & 0.13 & 0.07 & 0.13 & 0.94 & 0.93 & 0.07 & 0.07 & 0.26 & 0.50\\
18 & 0.08 & 0.38 & 0.95 & 0.91 & 0.15 & 0.23 & 0.03 & 0.16 & 0.02 & 0.13 & 0.95 & 0.92 & 0.05 & 0.07 & 0.09 & 0.48\\
19 & 0.03 & 0.08 & 0.95 & 0.94 & 0.10 & 0.14 & 0.01 & 0.03 & 0.01 & 0.03 & 0.95 & 0.94 & 0.04 & 0.05 & 0.03 & 0.10\\
20 & 0.01 & 0.08 & 0.94 & 0.94 & 0.06 & 0.15 & 0.01 & 0.03 & 0.00 & 0.03 & 0.94 & 0.95 & 0.02 & 0.05 & 0.01 & 0.10\\
\addlinespace
21 & 0.91 & 0.90 & 0.09 & 0.18 & 0.19 & 0.21 & 0.32 & 0.31 & 0.20 & 0.21 & 0.34 & 0.54 & 0.08 & 0.08 & 0.75 & 0.78\\
22 & 0.84 & 0.87 & 0.10 & 0.26 & 0.21 & 0.24 & 0.32 & 0.33 & 0.17 & 0.20 & 0.17 & 0.60 & 0.05 & 0.09 & 0.65 & 0.74\\
23 & 0.63 & 0.62 & 0.06 & 0.08 & 0.15 & 0.14 & 0.22 & 0.21 & 0.14 & 0.15 & 0.23 & 0.30 & 0.05 & 0.05 & 0.53 & 0.56\\
24 & 0.56 & 0.57 & 0.11 & 0.13 & 0.15 & 0.15 & 0.21 & 0.21 & 0.11 & 0.12 & 0.44 & 0.45 & 0.05 & 0.05 & 0.41 & 0.45\\
\addlinespace
25 & 0.19 & 0.36 & 0.95 & 0.91 & 0.20 & 0.21 & 0.07 & 0.13 & 0.07 & 0.13 & 0.95 & 0.92 & 0.07 & 0.07 & 0.25 & 0.50\\
26 & 0.08 & 0.38 & 0.97 & 0.91 & 0.15 & 0.23 & 0.03 & 0.15 & 0.02 & 0.13 & 0.97 & 0.92 & 0.05 & 0.07 & 0.09 & 0.48\\
27 & 0.03 & 0.08 & 0.97 & 0.95 & 0.10 & 0.14 & 0.01 & 0.03 & 0.01 & 0.03 & 0.97 & 0.95 & 0.04 & 0.05 & 0.03 & 0.10\\
28 & 0.01 & 0.08 & 0.96 & 0.94 & 0.06 & 0.15 & 0.00 & 0.03 & 0.00 & 0.03 & 0.96 & 0.95 & 0.02 & 0.05 & 0.01 & 0.10\\
\addlinespace
29 & 0.91 & 0.91 & 0.11 & 0.16 & 0.20 & 0.20 & 0.32 & 0.32 & 0.20 & 0.21 & 0.39 & 0.54 & 0.08 & 0.09 & 0.74 & 0.78\\
30 & 0.84 & 0.87 & 0.13 & 0.26 & 0.22 & 0.25 & 0.32 & 0.33 & 0.17 & 0.20 & 0.25 & 0.60 & 0.06 & 0.09 & 0.65 & 0.73\\
31 & 0.63 & 0.62 & 0.07 & 0.09 & 0.14 & 0.14 & 0.22 & 0.21 & 0.14 & 0.15 & 0.25 & 0.31 & 0.05 & 0.05 & 0.54 & 0.56\\
32 & 0.56 & 0.57 & 0.11 & 0.13 & 0.15 & 0.15 & 0.20 & 0.21 & 0.11 & 0.12 & 0.44 & 0.45 & 0.05 & 0.05 & 0.41 & 0.45\\
\bottomrule
\end{tabular}
\caption{Performance measures estimates for the hierarchical (Hier) and separate (Sep) models giving restricted mean survival time (RMST) and cure fraction ($\pi$) results. Values are averaged over all curves within scenarios.}
\label{tab:performance_measures}
\end{table}

\subsubsection{Performance of the target estimates}
For both the RMST and cure fraction the picture is similar. If the uncured survival curve and cure fraction prior distributions are informative then the separate and hierarchical models perform well with comparably small measures for bias, irrespective of whether an informative or weak prior is placed on the between group variance. However, in these scenarios the coverage is better for the separate model. Alternatively, when both uncured survival and cure fraction have only weakly informative priors then both separate and hierarchical models demonstrate poor bias, irrespective of the between group prior (scenarios 1-4, 9-12). The scenarios of note are when we have an informative prior for the uncured survival curve parameters but only weak prior knowledge for the cure fraction (scenarios 17-20, 25-28). In this case, the hierarchical model bias is clearly better than for the separate model.
The coverage is consistently better for scenarios with an informative survival curve prior too.
In term of sample size, the biggest difference between small and large samples in for when the cure fraction has a weak prior; or alternatively, with strong prior evidence then the measures are less sensitive to sample size. For coverage, when the bias is relatively large the coverage is low and vice-versa. In particular, for a small number of end-points and small sample size, the hierarchical model performs better in terms of bias when an informative survival curve prior is used with a weak cure fraction prior (scenarios 18 and 26). Conversely, for a weak prior for uncured survival and informative cure fraction (scenarios 5-8, 13-16) the coverage and bias are worse for the hierarchical model than the separate model, but in general when the uncured survival prior is weak the coverage is poor.

In summary, the pattern of the performance measures between target estimates is similar. This is likely because the RMST is  dependent on the cure fraction of the underlying curve. Overall, the simulation study shows that performance of estimates can vary depending on the cure fraction, uncured survival curves and the sample size. The scenarios when there is less prior knowledge about the uncured survival curve and more information about the cure fraction seem to preference the separate model, whereas conversely scenarios where the hierarchical model is most beneficial include those with informative priors on the uncured survival and uncertainty about the cure fraction. This latter case seems reasonable in reality when, for instance, a new drug impacts an uncertain proportion of long-term survivors for a well understood underlying process. In contrast, in other situations, it may not be possible to obtain direct evidence about an untreated population due to ethical considerations.

\section{Real data example}\label{sec:application}
In this section we apply the models from Section~\ref{sec:hier_model} to the CheckMate 067 dataset. We first detail the modelling aspects specific to this analysis before presenting the results.

% software
This analysis was carried-out using the Stan inference engine \cite{carpenter2017stan} called from R v.4.3.1~\cite{Rcoreteam} using the package \texttt{rstan} on a Windows 11 PC with 12 cores. Each fit employed the default of 4 chains of 2000 iterations each, and a warm up of 100 iterations. In order to determine convergence, we checked effective sample sizes, Markov chain standard error and performed posterior predictive checks. Details of the algorithm and model checking can be found in the Appendix. The code has been developed into an R package using a generalisable framework. This means that analyses for an arbitrary number of end-points can be easily performed with new data. The package is freely-available and can be downloaded from \url{https://github.com/StatisticsHealthEconomics/multimcm}.

\subsection{Background survival}
In many cure fraction models, it is common to have $S_b(t, x) = 1$ for all $t$ under consideration, which is reasonable in the short-term. However, we are focused on the full life-course of an individual in an HTA context and so include a general population all-cause mortality competing risk. Thus, the uncured population is subject to a background mortality risk in addition to the (excess) risk from the event of interest.

We used the World Health Organisation (WHO) life tables by country for the latest year available of 2016 \cite{wholifetables} to inform the background mortality in the mixture cure model. The baseline hazards are the expected mortality rate for each patient at the age at which they experience the event. The mortality data are adjusted for country, age and gender. Thus, this structure provides a granular account of the different patient profiles in the trial. The WHO reports conditional probabilities of death in 5-year intervals until age 85. A constant annual mortality rate is reported for individuals over 85. The life table is produced assuming that no-one lives beyond 100 years. Further details are provided in the Appendix.

In a Bayesian analysis, there are alternative ways in which we could model the background mortality. For this work we used WHO hazard point estimates as fixed and known. We could consider the WHO estimates to provide sufficiently accurate estimates given the sample size, and so incorporating uncertainty is less crucial than in other cases. This also forces consistency across fits. Denote the fixed WHO estimates for individual $i$ used for the background survival in equation (\ref{eqn:mcm}) as $S_b(t_i \mid \bm\theta_b, \bm{x}_i) = \hat{S}_i$.

% background adjustment
It may be that the patients in the trial have a worse background survival than the average equivalent individual in the general population, and so naively using the WHO life tables will over-estimate their survival times. There are several possible ways in which this could be addressed. For example, using only the complete responders in the sample, who are clinically confirmed as cured, it is possible to obtain a posterior distribution for a hazard ratio between these and a WHO estimate baseline. This can serve as a prior in the main model in a two-step approach. Alternatively, prior belief could be defined explicitly using expert knowledge. This could be elicited directly for the hazard ratio or on a natural scale, such as mean lifetime, and transformed.

\subsection{Prior specification}
For the latency model, we centred covariates to aid computation and specified minimally informative priors on the log-scale for the coefficients of the OS and PFS rates, $\log(\lambda_{OS})$ and $\log(\lambda_{PFS})$, respectively. These were the same for both OS and PFS: ${\beta_{\cdot,0}^{\lambda} \sim \text{Normal}(-3, 0.5)}$ and for age $\beta_{\cdot,age}^{\lambda} \sim \text{Normal}(0, 0.01)$. These values were picked to be vague about the true rate but the median survival time was expected to have been reached before 30 months. Figure~A4 in the Appendix shows realisations of survival curves using samples from the prior for $\beta_{\cdot,0}^{\lambda}$. At 30 months the median survival probability is 0.22 [IQR: 0.01 - 0.58] and at 60 months is 0.05 [IQR: 0 - 0.34]. The prior distributions for the ancillary parameters $\phi_{OS}$ and $\phi_{PFS}$ depended on the particular survival distribution used. For log logistic and Weibull the prior distribution for the shape parameter was $\phi \sim \text{Gamma}(1,1)$, for Gompertz $\phi \sim \text{Gamma}(1,1000)$, and for log Normal the prior distribution for the standard deviation was $\phi \sim \text{Gamma}(1,2)$.

For the incidence model, the global cure fraction was modelled as a fixed effect linear regression with logistic link. Each treatment coefficient prior is independent $\beta^{\pi}_k \sim \text{Normal}(-0.1, 0.2)$. This corresponds to a prior mean cure fraction of just under 0.5, and approximately 10\% and 1\% chance of exceeding 0.6 and 0.7, respectively. This is a relatively weak prior (see simulation study) but still informed by expert judgement. Alternatively, we could have specified the cure fraction on the natural scale between 0 and 1 using a Beta prior distribution $\pi \sim \text{Beta}(a, b)$. This has the advantage of representing the uncertainty about the parameter directly which could aid elicitation but has the disadvantage of $a$ and $b$ being less interpretable. The Beta distribution parameters can always be obtained via transformation of the mean and standard deviation of the Normal distribution. Prior predictive distribution plots and further discussion are provided in the Appendix.
%GB: This section is generally confusing. I think you're trying to be too quick and the reader may struggle to navigate what you're doing... It took me a couple of passes to understand what you meant by \pi as a Beta or \pi as a regression. And the language could be tightened up a bit..

\subsubsection{Cure fraction between-endpoint variance}
% complete data
For the complete data analysis, the random effect variance on the global cure fraction was modelled using a minimally informative (folded) half-Normal \cite{Gelman2006} with  ${\sigma_k \sim \text{Normal}(0, 2.5^2)\mathbb{I}(0,)}$, where $\mathbb{I}(0,)$ denotes a distribution truncated at mean 0. We considered several alternatives following the recommendations in Gelman~(2006) and this was found to be the best option for this study because it balances vagueness and support at zero. As a rule of thumb, when the number of groups is greater than 5 then a minimally informative uniform prior density associated with the standard deviation $\sigma$ is expected to generally work well, or when there is small hierarchical variance then a $\text{Gamma}(2, 0.1)$ may be preferred \cite{Chung2013}.

% cut-point data
As discussed, the early phase of a trial is not very informative of the cure fraction if plateauing is yet to occur. In this situation, we can take advantage of the Bayesian hierarchical structure. In the absence of other evidence, it is reasonable to assume that uncertainty about the cure fractions for each end-point can be modelled using a common distribution, which corresponds to a narrow prior on the between-group standard deviation at the global cure fraction level.
In order to do this, we modelled the global cure fraction with a {\it Penalised Complexity (PC)} prior \cite{Simpson2017} that penalizes departure from a base model and preferences simplicity.
That is, the base model as the pooled data model is the most parsimonious model. 

%% truncated Normal prior
% Using the same prior distributions as the main complete data analysis, then the global cure fraction was modelled using a two-tailed confidence interval of approximately 99\% of the density so significance level $\alpha = 0.01$. This corresponds to a truncated Normal distribution $\sigma_k \sim \mbox{Normal}(0, 0.06^2) \mathbb{I}(0,)$ for $p(\pi_k > 0.55) = 0.005$, ie, 99\% interval [0.45, 0.55], and $\sigma_k \sim \mbox{Normal}(0, 0.03^2)\mathbb{I}(0,)$ for $p(\pi_k > 0.525) = 0.005$.

\begin{figure}[!ht]
\centering
\includegraphics[height=7cm, width=0.9\linewidth]{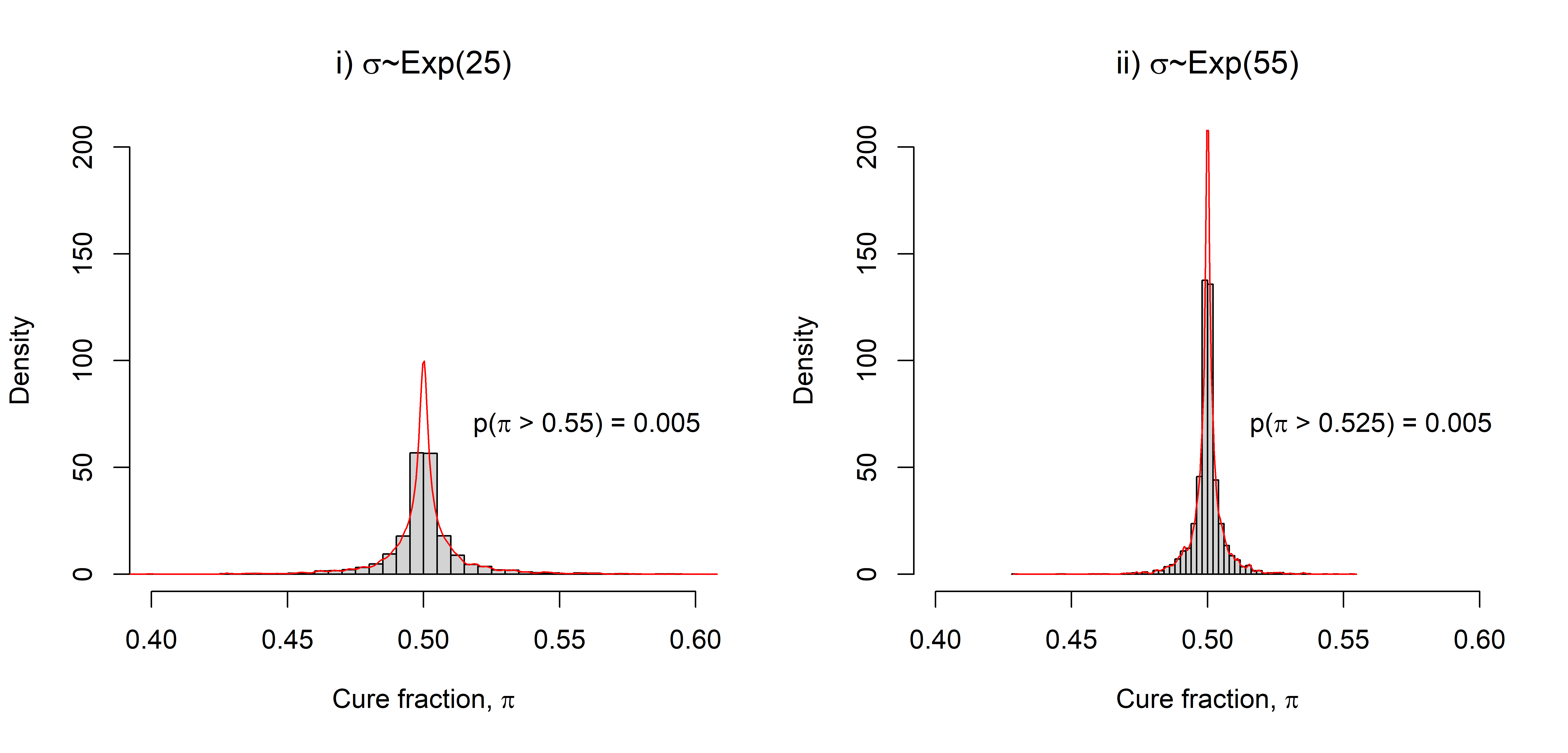}
\caption{\label{fig:complexityprior} Penalised complexity prior distributions on the cure fraction scale corresponding to different between-group variances at the global cure fraction level of the Bayesian hierarchical mixture cure model.}
\end{figure}

In practice, the PC prior approach places an exponential distribution on $\sigma_k$ in equation (\ref{eqn:global_cf}).
The rate parameter for this distribution, $\rho$, is obtained by first specifying a threshold value $\sigma_0$ and probability of exceeding that threshold, $\alpha$, ie, $p(\sigma > \sigma_0) = \alpha$. These can then be plugged into $\rho = - \log(\alpha)/\sigma_0$, to obtain the rate $\rho$ of the exponential distribution prior.
For specific values, for $p(\pi_k > 0.55) = 0.005$ then $\sigma_0 = 0.18$ and $\rho = - \log(0.01)/0.18 = 25$ or $\sigma \sim \mbox{Exp}(25)$. Similarly, for $p(\pi_k > 0.525) = 0.005$ then $\sigma_0 = 0.08$ and $\rho = - \log(0.01)/0.08 = 55$ or $\sigma \sim \mbox{Exp}(55)$.
The latter prior was used for the data-cut at 12 months. When more data are available at later times the use of this prior was replaced by the truncated half-Normal as discussed elsewhere. This approach is in the spirit of information-sharing methods such as power priors, which are "adaptively robust", where a power parameter controls the degree of borrowing. For example, for the simple case of normal historical data under a flat prior, the power prior is another normal distribution with a scaled variance according to the sample size and magnitude of the power parameter, determined by the investigator.
We can see how the two prior distributions correspond to the prior uncertainty on the cure fraction scale in Figure~\ref{fig:complexityprior}. The two distributions are similar, but the plot for $\rho = 55$ is more peaked around the central value, preferencing the base model even more.
 
\subsection{Parametric models for the sampling variability}
To model the sampling variability in the observed OS and PFS event times we used the exponential, Weibull, Gompertz, log-normal and log-logistic distributions. These are among those recommended by NICE guidelines for Health Technology Assessments (HTA) \cite{Latimer2011}.
In general, providing this list has been interpreted by many in the HTA literature as meaning all of these distributions should be modelled with a given data set. This prescriptive approach does not take into account what may be known {\it a priori} about the problem. In reality, a subset of these parametric models will be more plausible for the data and so these are the ones that should be investigated.

\subsection{Model assessment}
We evaluated the goodness of fit of the overall models using out-of-sample predictions estimated with the widely applicable information criterion (WAIC) (and leave-one-out cross-validation, LOO, given in the Appendix) \cite{Vehtari2017}. These have various advantages to more common AIC and DIC. WAIC is the difference between the computed log pointwise predictive density and the estimated effective number of parameters. Smaller values are better. It is fully Bayesian and invariant to parameterisation. They are easily obtained using the posterior sample of the Stan output.
% The effective number of parameters is not uniquely defined because it depends on the level of the hierarchy that is in {it focus} \cite{spiegelhalter}.
%
% The set of parameters in the hierarchical model (and their dimensions) are
% $\mathbf{\beta^{\pi}}$ (3), $\mathbf{\sigma}$ (3), $\mathbf{\beta^{\lambda}_{PFS}}$ (2), $\mathbf{\beta^{\lambda}_{OS}}$ (2), $\mathbf{\phi_{OS}}$ (1), $\mathbf{\phi_{PFS}}$ (1), $\mathbf{\pi_{OS}}$ (3), $\mathbf{\pi_{PFS}}$ (3).
% Depending on the number of ancillary parameters the total is 16, 17 or 18.

% {\it We also calculated the 
% $\Delta S$ at months $t = 10, 20, 30, ...$\\
% clinical significant vs statistical significant differences.
% hazard ratio?\\
% median/mean survival?}\\

%%TODO: include this?
%% CrI widths
% In order to investigate potential decrease in the degree of uncertainty of the estimates as a result of borrowing of information across end-points, we calculated ratios of the width of the 95\% CrIs.
% Define as the ratio of the widths of the CrIs of $\pi_{os}$ and $\pi_{pfs}$ from the hierarchical model to the width of the CrIs of the separate model.

There has not been much work carried-out to date on model assessment methods specific to mixture cure models. For the basic mixture cure model, assessment may focus on the model as a whole or on the incidence and latency submodels separately. The fit of the latency submodel can be examined using a revised Schoenfeld residual \cite{Wileyto2013}, and the performance of the incidence submodel can be checked with concordance measures such as AUC \cite{Peng2021}. However, these incidence model approaches are best used directly with known cure status and are designed for models which consider individual cure fractions, unlike our model which assumed a single population cure fraction. Of course, we also extend the basic mixture cure model so it is not yet clear how to best use these methods in this situation. 

\subsection{Results using the complete data set at 60 months} \label{sec:results}
In this section, we will explore the mixture cure model fits using the complete 60-months CheckMate 067 trial data. An example of the Bayesian hierarchical model posterior survival curves using the complete data and exponential distributions for both OS and PFS is shown in Figure~\ref{fig:S_exp^{cens}f_hier}. We show this because, even though the exponential is the simplest of the NICE recommend distributions, from visual inspection it fits fairly well. Equivalent plots for all other distributions and a combined plot are given in the Appendix.

The background survival appears to decrease at a small, steady rate and is the same between plots, as expected because the point estimates from WHO life tables were used directly. Visual inspection demonstrates reasonably good model fits. The PFS Kaplan-Meier curves show an initial steep decrease before levelling off. The exponential fits particularly well to the tail which is of particular importance because this is usually where extrapolations have the worst fit.

% restricted mean survival times
For a follow-up time of 60 months and for the exponential model, the RMSTs (95\% credible interval, CrI) for the uncured group are 20.6 (18.4, 23.1) and 6.66 (6.09, 7.39) months, for OS and PFS, respectively. For the OS end-point, the RMSTs are 28.7 (26.3, 31.1), 36.5 (34.1, 38.9) and 39.6 (37.1, 42.1) months, for ipilimumab, nivolumab and combined, respectively. For the PFS end-point, the RMSTs are 11.9 (10.0, 14.1), 24.3 (21.1, 27.4) and 27.7 (24.7, 30.8) months, for ipilimumab, nivolumab and combined, respectively. The RMST for the cured population is the same value corresponding to the general population of 58.2 months.

% % time to reach 0
% The latest time PFS has survival probability greater than 0 is over 30 months for the exponential distribution.
% Similarly, the OS survival probability is greater than 0 at the end of follow-up at 60 months.

%%TODO: include a table of survival statistics?
%%      compare curves formally? e.g. ANOVA, t-test RMST or HR?

\begin{figure}[!ht]
\centering
% produced using paper_output_plots.R
\includegraphics[width=0.7\linewidth]{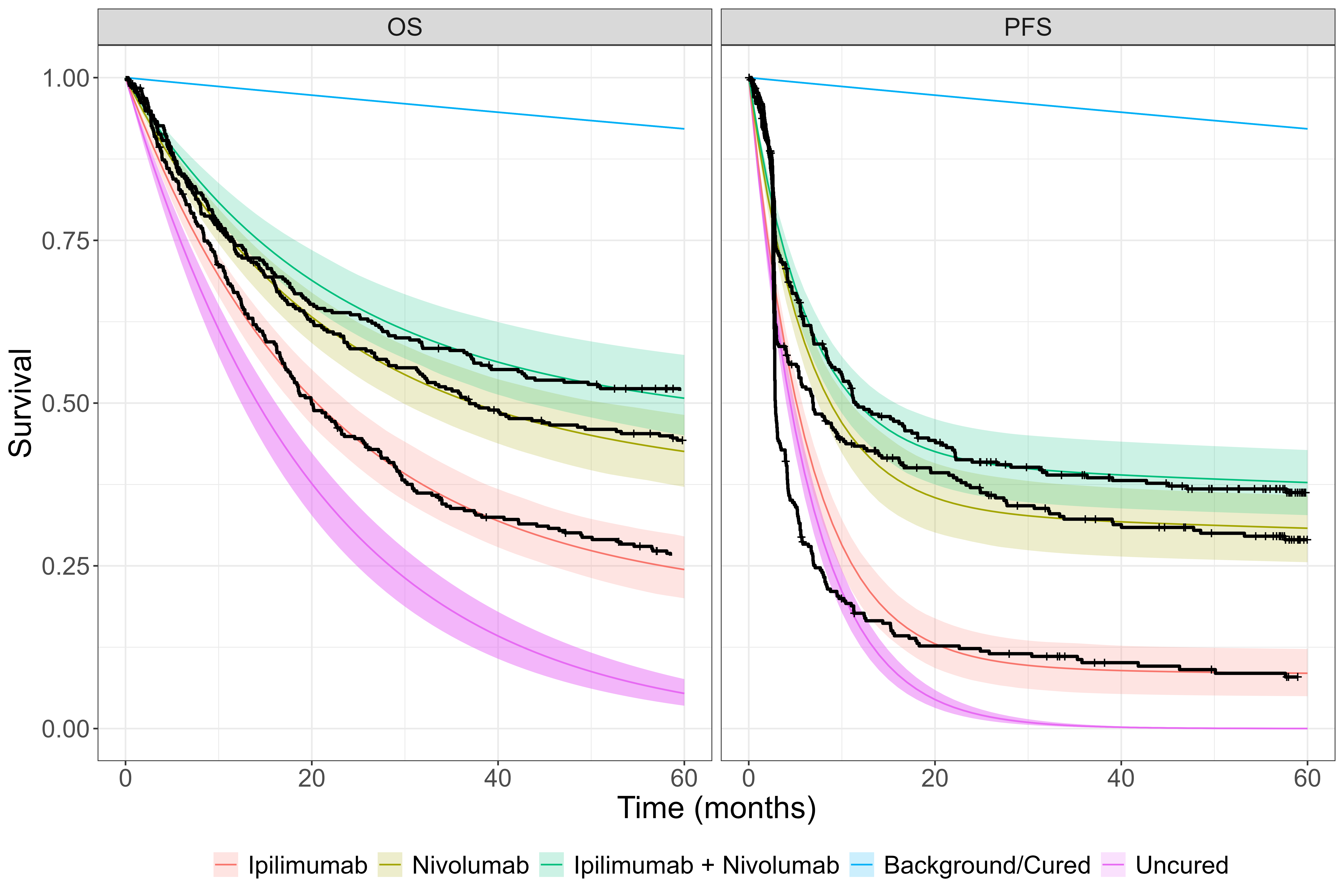}
\caption{\label{fig:S_exp^{cens}f_hier} Bayesian hierarchical mixture cure model posterior mean survival curves for uncured fraction assumed exponential for both OS and PFS, and ipilimumab, nivolumab and combination treatments. The black lines show the Kaplan-Meier curves using the complete CheckMate 067 trial data.}
\end{figure}

% cure fractions

% Figure~\ref{fig:cf_forest_all_tx} and Figure~\ref{fig:cf_forest_all_tx_sep} show cure fraction posterior distribution forest plots for the hierarchical and separate mixture cure models, respectively. 

% global cf
The global cure fraction posterior is relatively wide, which is understandable since there are only two exchangeable points providing limited information (OS and PFS). They range 0.1-0.55, 0.2-0.6, 0.2-0.65 for ipilimumab, nivolumab and combined treatments respectively. However, despite this, the data influence the end-point cure fractions by pulling the central location away from the prior, downwards for ipilimumab and upwards for the combined treatment. For ipilimumab, the PFS curve fraction posterior is close to containing zero within the credible interval. Whereas for nivolumab and the combined treatment arms the mean global cure fraction lies somewhere between the PFS and OS means, the ipilimumab global mean is closer to the OS cure fraction mean. This is because we have explicitly codified in the prior distribution that the global cure fraction is unlikely to be near zero and the PFS cure fraction posterior is close to zero. We would observe the equivalent behaviour if the OS distributions were near to 1.
% plateau time
The survival curves for the whole sample plateau to the background mortality survival from when the uncured survival probability reaches zero. Since the background is user-supplied, in this case from the WHO life tables, then there is no additional insight from this time on-wards. The times at which this occurs are relatively early for PFS and the log-Normal distribution.
% hier vs separate
Comparing the hierarchical and separate models there is not a noticeable difference between the two. As mentioned above, there is limited information with only two end-points and a fairly weak prior. There is a small amount of shrinkage such that the posterior distributions for OS and PFS are pulled toward the global mean, so that the PFS cure fractions are larger and the OS cure fractions are smaller in the hierarchical model than the separate model.

\subsection{Results using all data-cuts}\label{sec:results-data-cut}
We now turn our attention to fitting the mixture cure models to earlier data-cuts in the CheckMate 067 trial data set.
For the complete data set up to 60 months we have seen that the separate and hierarchical models produce similar parameter inferences. This is to be expected when the data are relatively mature, there are in effect only two data points at the global cure fraction level and a weak prior is used for the global variance hyper parameter. For demonstration purposes, we show results using the penalised complexity prior with exponential distribution rate parameter $\lambda = 55$ for the 12 months data cut scenario and retain all other previous priors, including the weak priors for the mean $\pi_k$. Recall that for later times, when more data are available, the alternative variance prior was adopted.

\begin{figure}[!ht]
\centering
\subfloat[Separate Mixture Cure Models]{
    \includegraphics[trim=0 0 220 0, clip, height=10cm, width=0.45\linewidth]{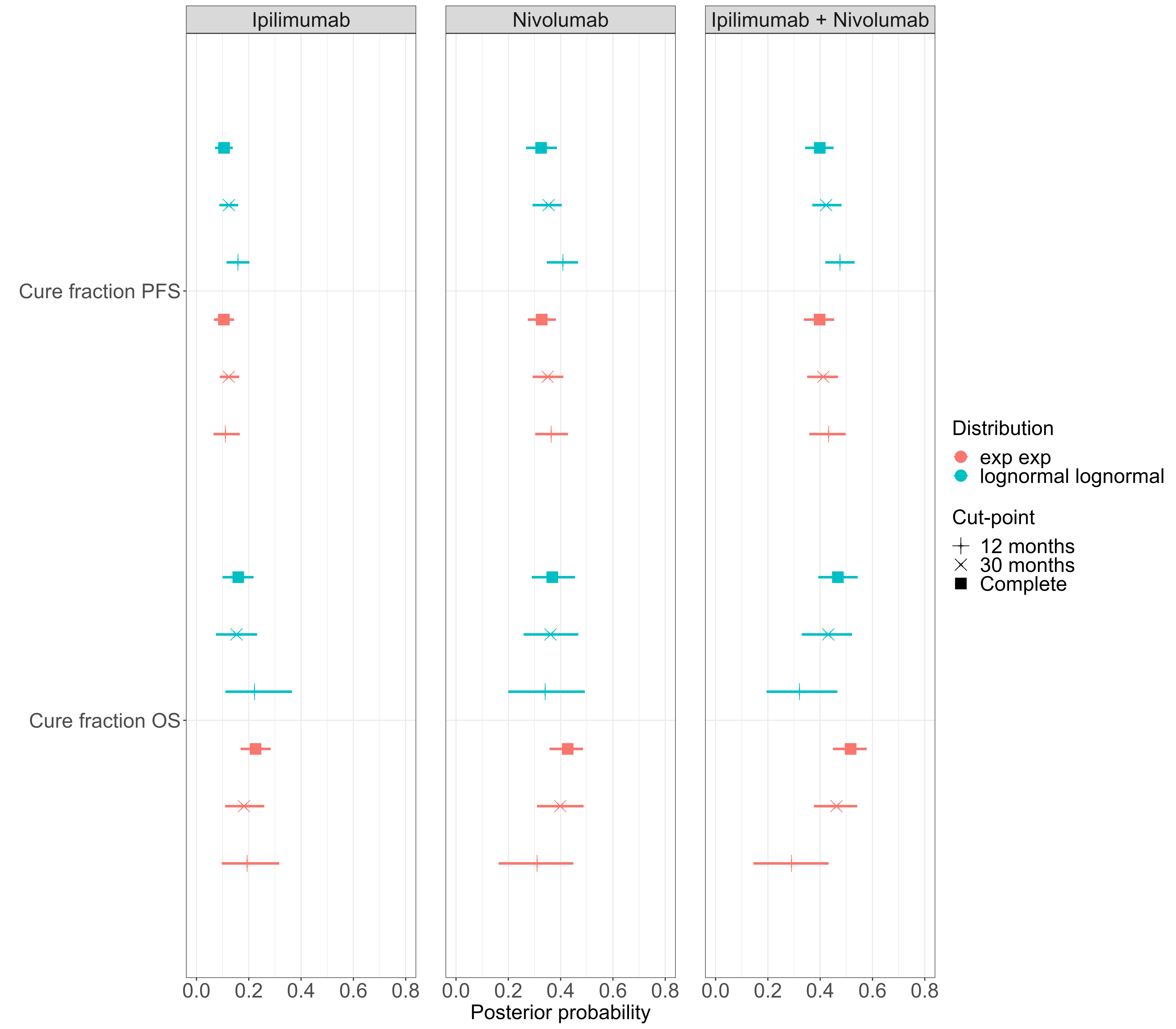}
    \label{fig:forest_plot_cf_cutpoint_sep}
}
\subfloat[Hierarchical Mixture Cure Model]{
    \includegraphics[height=10cm, width=0.55\linewidth]{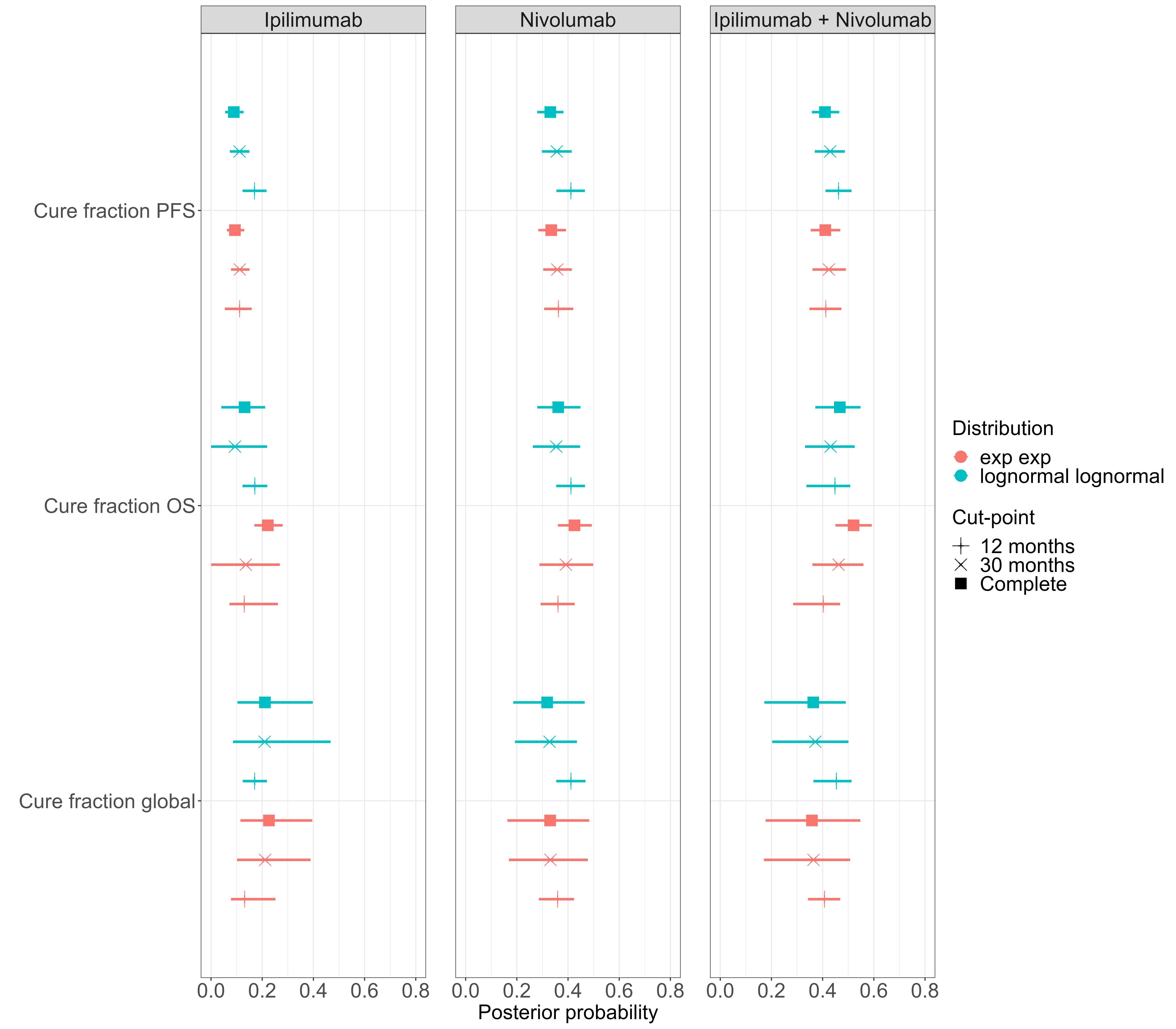}
    \label{fig:forest_plot_cf_cutpoint_hier}
}
\caption{Posterior cure fraction forest plots with 95\% credible intervals for study data with cutpoint censoring at 12 months, 30 months, and the complete dataset.}
\label{fig:combined_forest_plots}
\end{figure}

Figures~\ref{fig:forest_plot_cf_cutpoint_sep} and Figure~\ref{fig:forest_plot_cf_cutpoint_hier} show forest plots of the cure fraction estimates for 12, 30 and 60 months, and for the separate and hierarchical models, respectively.
Results are shown for the OS exponential with PFS exponential model and the OS log-Normal with PFS log-Normal model.
We can see that in Figure~\ref{fig:forest_plot_cf_cutpoint_sep} for the exponential model in the OS survival curves for the separate model there is more uncertainty than in the equivalent plot for the hierarchical model in Figure~\ref{fig:forest_plot_cf_cutpoint_hier}. The log-Normal model OS survival curves for separate and hierarchical models over-estimate the cure fraction at earlier data-cuts and all treatment arms. For later data-cuts, the cure fractions appear to converge from above to the complete data estimates.
% It appears to revert to the background survival at the time of the last observed data point.

Conversely, for the separate models and all treatment arms, at 12 months follow-up the exponential model underestimates the OS cure fraction and approaches the complete data value from below. This is less clear than for the log-Normal model probably because of the larger estimate uncertainty for the exponential model. Crucially, these are two clearly different behaviours but we would be ignorant to which is correct (if any) until later on in the trial. For OS in the hierarchical model, the cure fraction estimates appear to be accurate and more certain even at 12 months for all treatment arms. The posterior is more stable about the complete data cure fraction for all considered data-cuts.

\begin{figure}[!ht]
\centering
\includegraphics[height=10cm, width=0.6\linewidth]{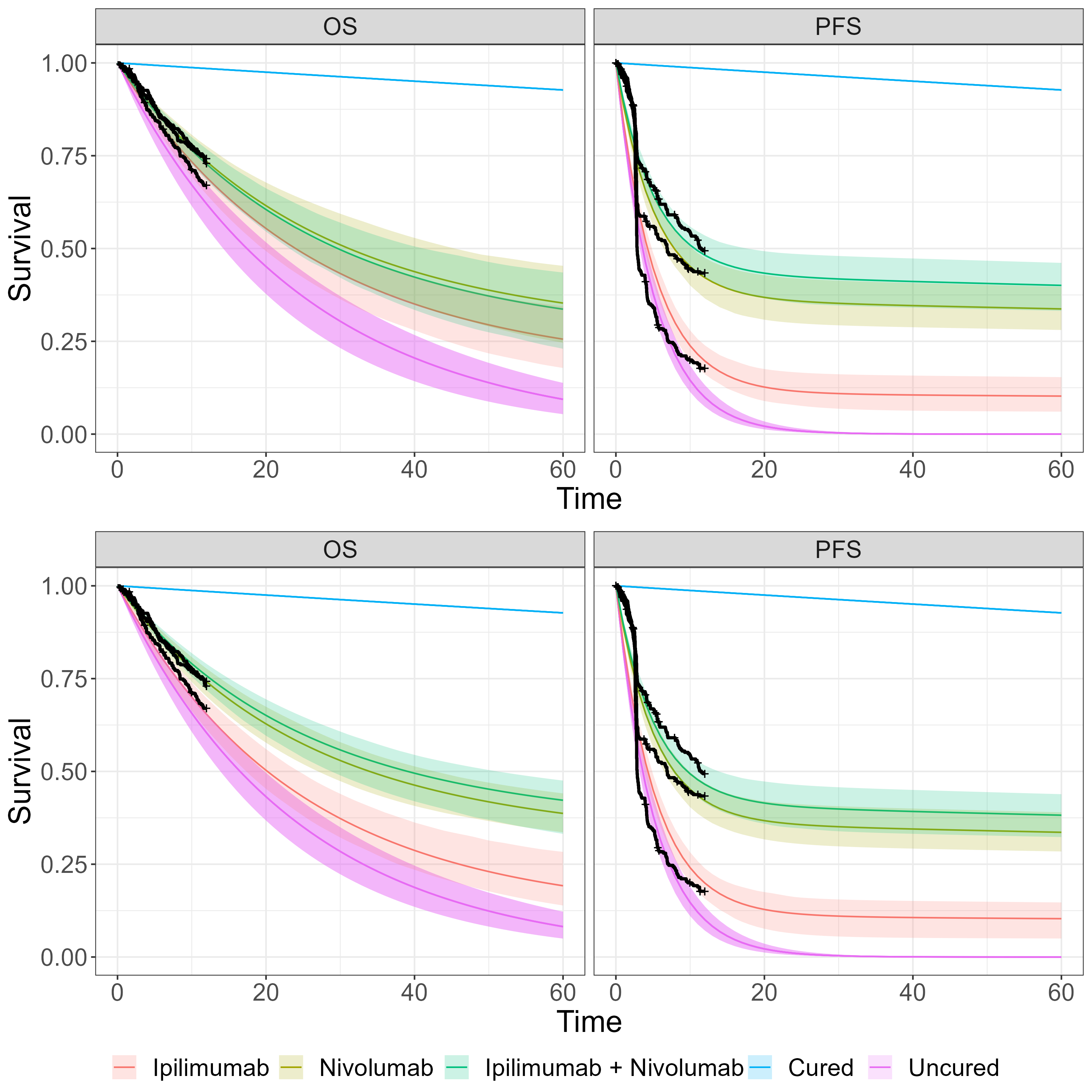}
\caption{\label{fig:S_cutpoint_12mo_exp} Mixture cure models posterior survival curves with 95\% credible intervals using cut-point data for 12 month follow-up and for separate (top) and hierarchical (bottom) models. Both PFS and OS use exponential uncured survival curves. The black lines show the observed data Kaplan-Meier curves.}
\end{figure}

Figure~\ref{fig:S_cutpoint_12mo_exp} shows the survival curves for the data-cut with 12 months of follow-up data for the exponential and for the separate and hierarchical models, respectively. Figures for 30 months data-cut and table of the survival estimates at 60 months extrapolated using 12 and 30 months data cut point are given in the Appendix.

%% restricted mean survival times
Comparing the uncured RMSTs for various models,
the hierarchical exponential model gives for OS 20.6 (18.4, 23.1) and PFS 6.72 (6.15, 7.34) months. For the separate model, OS is 22.3 (18.5, 26.9) and PFS is 5.68 (4.9, 6.45) months. For the log-Normal hierarchical model, the RMST for OS end-point is 6.6 (6.17, 7.13) and PFS is 3.98 (3.84, 4.12) months. Finally, the separate model gives 6.59 (6.18, 7.07) months for OS and 3.98 (3.85, 4.1) months for PFS.

% \clearpage  % force figures before section
%
\section{Discussion}\label{sec:discussion}
We have presented a Bayesian hierarchical mixture cure model to obtain complete survival curves that exhibit plateau behaviour. These curves can then be used in HTA to inform functions of patient lifetime. We also conducted a simulation study demonstrating situations where the hierarchical or separate models are more appropriate. Finally, we demonstrated the novel method using the CheckMate 067 trial data set for end-points OS and PFS and at different artificial data-cuts; incorporating additional structure for this particular situation. The hierarchical model was compared with the separate model analogue and has been shown to be superior, as well as both principled and applicable to situations where trials include multiple treatment arms and end-points.

% Bayesian benefits
Our approach may be preferable to a frequentist or non-hierarchical model both practically, as shown in the simulation study, and conceptually, given the theoretical framework underpinning it. A benefit to adopting a Bayesian paradigm over a frequentist approach is the flexibility in modelling complex structure, in our particular case multi-level or nested data. In contrast, a frequentist model would need to impose assumptions about the structure of a covariance matrix. Another advantage of adopting a hierarchical structure in a Bayesian approach is robustness to smaller sample sizes due to the provision of prior information and borrowing of strength between data for different end-points, leading to more stable, accurate extrapolation estimates. Principled prior distribution specification about the cure fractions was shown to be especially useful early on in the trial when observable data about long-term survivors was weak. This results in the stabilisation of the cure fraction estimates, which again results in improved extrapolation. In general, extrapolations of survival outcomes with a shared cure fraction provide a more conservative and practical approach to modelling the end-points separately. It may be that there is excessive borrowing of information between cure fractions when the end-points are distinctly different. In our example, this was not necessary, since there are only two end-points but for a larger number this may be appropriate. Partial-exchangeability can be used instead in this case \cite{Neuenschwander2016}.

% choice of priors
In comparison to some previous frequentist approaches, our framework is able to include contextual information about the parameters of the model via the prior distributions. Incorporating sensible information, both in the form of external data sources or elicited from experts can stabilise the inference on the cure fraction and restrict/constrain its values via the prior distributions. For the analysis presented in this paper, we exploited different priors. Minimally informative priors allow the data to dominate the posterior, even when the dataset is relatively small. In practice, there may be additional information from clinical experience or previous data that can better inform the prior selection. The priors can be specified on the natural scale to aid elicitation and interpretation. For example, we may have clinical knowledge that a drug treatment could generate a cure fraction greater than, say, 30\%. If a (cancer) patient enters the study at 60 years old then there is a high degree of certainty that they will not live for, say, over 40 years. In the case of sparse data e.g. when there is a large amount of missing data perhaps due to right censoring, then `regularising' the inference based on available prior information could make a significant and important difference. In particular, the method of blending curves is a recent, straightforward option for how to constrain survival \citep{Che2022}. How to elicit this prior information in practice may not be simple and a formal protocol should be adopted \cite{OHagan2019}.

Healthcare regulators, such as the European Medicines Agency (EMA) and U.S. Food and Drug Administration (FDA), may value the fact that a hierarchical model can embed some form of prior knowledge to account for scepticism. This will prevent the cure rates to be taken at face value, particularly with limited data. The hierarchical model seems to give a more precise estimate of the cure rate, even with earlier data-cuts (especially for the exponential model). The hierarchical model is more aligned, which means the possibility of more reliable estimates early on, which is a good argument to complement modelling based on earlier cuts.
% Earlier cuts can lead to quicker introduction in the market and reimbursement decisions.

% limitations
In our analysis, we simulated censoring due to data-cuts by simply censoring all patients in the complete data set at the same time point. In practice, a more realistic censoring time might have been the date of last visit for each patient; however, this would not have affected the results of this analysis. Primarily, the data-cut points were selected to demonstrate the theoretical justification of our method and not to strictly follow the study data-cut protocol.

% data
Future work should test the proposed hierarchical MCM on a variety of data sets and at data-cuts reflecting their actual trial protocols. The modelling framework is generalisable from two end-points in our example to any number in principle. For example, previous related frequentist analyses considered clusters of hospitals and recurrent events per individual \citep{Lai2008, Lai2009} or interspecies extrapolation in dose-response experiments \citep{DuMouchel1983}.
Further work could investigate the benefit to including additional end-point data, even if it does not clearly demonstrate plateauing behaviour. The background survival curve used in the model representing the statistically `cured' survival in general population can be refined to be appropriate to specific cohort characteristics. For instance, long-term survival of clinically identified complete responders (CR) could be used to bridge the gap between statistical cure and clinical cure/functional cure. The shorter-term CR individual-level survival data can be combined with longer-term external data, such as from WHO life tables as in the current analysis (see \citep{Jackson2017} for potential approaches). As a way to model the constraint that $S_{OS} \geq S_{PFS}$ we could consider an excess hazard-type formulation such as $h_{OS} = h_B + h_{OSe}$ and $h_{PFS} = h_B + h_{OSe} + h_{PFSe}$, where $h_{OS}$ is the hazard for all cause mortality, $h_{PFS}$ is the hazards for either progression or all cause mortality, $h_B$ is the background hazard, $h_{OSe}$ is the excess hazard for overall survival and $h_{PFSe}$ is the additional excess survival due to progression alone.

% optimal stopping
Another way to think about the cut-point data is to reflect on what is the minimum follow-up time should we choose to halt the trial and end the data collection which would maximise some expected overall utility.
% RSS conference talk ref? Anna?
That is, put another way, what is the optimal stopping time $\tau^*$ for the total utility $V$?
$$
\mathbb{E} V_{\tau^*} = \sup_{\tau} \mathbb{E} V_{\tau}
$$
The stopping utility should incorporate a measure of the survival extrapolation robustness and may include benefits, such as smaller costs, freed-up time from the trial for patients and those running the study, quicker time to market for a drug, and avoidance of side-effects.
The utility would also depend on estimating `correct' parameters values from the available data ie, that they are within some acceptable threshold of the true values.
This could be applied directly on the parameters of interest or an order statistics for which drug is deemed preferable.

This may be framed in terms of value of information (VoI) \citep{Heath2017}. For example, in the case given above, what is the additional value of obtaining further data after 12 and 30 months. In practice, deriving a set of heuristics may be better than formalising a general rule. A discrete set of cut-points may be used, similar to our artificial selection of rounded times 12 and 30 months, rather than considering the whole real line.

As well as during a study, the hierarchical MCM could be used in advance of a study to inform trial design and optimal follow-up time to support the choice of data-cuts used in practice. Smaller sample sizes and shorter duration may provide equivalent power given the exploitation of additional model structure.

% implications for HTA
There are several implications for HTA by using the hierarchical mixture cure model presented here. In a cost-effectiveness analysis, unstable estimates over time would give different estimates of benefit and thus cost-effectiveness statistics and potentially differing optimal decisions. Lee~(2019)\cite{Lee2019} performed a cost-effectiveness analysis of using nivolumab with ipilimumab vs ipilimumab using the CheckMate 067 trial data, and a partitioned survival model and Markov state-transition model. Lifetime costs and benefits were estimated. Results were obtained using 18-month (when OS data were unavailable) and 36-month (OS available) CheckMate 067 data-cuts. They showed that the model using OS data generated more than 1 additional quality-adjusted life-year (QALY) across both treatment arms compared to the model without OS data. Using our approach, we have shown that reliable estimates of complete survival curves for OS can be obtained even for early data-cuts. These can be used to estimate lifetime horizon benefits in cases where such estimates were previously unavailable. In our example, the OS cure fraction is accurate and stable at 12 months for the hierarchical mixture cure model with exponential distributions for the OS and PFS uncured survival models. Additional benefits are potential early end of trial (cheaper, improved health, quicker drugs to market for producers and patients benefit), and smaller trials.

%\backmatter

%%TODO:
% \section*{Acknowledgements}
% We thank Victoria Federico Paly ...

% \subsection*{Author contributions}

% This is an author contribution text.

% \subsection*{Financial disclosure}

% None reported.

% \subsection*{Conflict of interest}

% The authors declare no potential conflict of interests.

% \section*{Supporting information}

% The following supporting information is available as part of the online article:

% \newpage
\clearpage

% Stan?
% \begin{lstlisting}[caption = {Descriptive Caption Text}, label=DescriptiveLabel]
% for i:=maxint to 0 do
% \end{lstlisting}

% %\nocite{*}% Show all bib entries - both cited and uncited; comment this line for only cited entries

%% for arXiv submission

% \clearpage

% \section*{Author Biography}

% \begin{biography}{\includegraphics[width=66pt,height=86pt,draft]{empty}}{\textbf{Author Name.} This is sample author is sample author biography text.}
% \end{biography}

\end{document}